# On the mechanism of gas adsorption for pristine, defective and functionalized graphene

Y. You[1], J. Deng[1], X. Tan[2], N. Gorjizadeh[1], M. Yoshimura[3], S. C. Smith[2], V. Sahajwalla[1], R. K. Joshi[1]

[1]Centre for Sustainable Materials Research and Technology (SMaRT), School of Materials Science and Engineering, University of New South Wales Sydney, Australia
[2] Integrated Materials Design Centre (IMDC), School of Chemical Engineering, University of New South Wales Sydney, Australia
[3]Surface Science Laboratory, Toyota Technological Institute Nagoya 468-8511, Japan

*Defect is no longer deemed an adverse aspect of graphene. Contrarily, it can pave ways of extending applicability of graphene. Here, we discuss the effects of three types of defects on graphene: carbon deficiency, adatom (single Fe) dopant and introduction of functional groups (carboxyl, pyran group) on $NO_2$ gas adsorption via density functional theory method. We have observed that the unsaturated carbon in defected graphene is highly active to attract $NO_2$ molecules. Our study suggests that introducing Fe on graphene can enhance the $NO_2$ adsorption process. Adsorption energy calculations suggest the enhancement in $NO_2$ adsorption is more profound for Fe-doped mono and tetra vacant graphene than Fe doped bi- and tri-vacant graphene. This study could potentially be useful in developing adsorption-based applications of graphene.*

Graphene is regarded as a promising material for many practical applications[1–7]. When used for gas sensing applications, the extremely high active surface area and fast carrier mobility endow graphene the ability to detect a single molecule of target gas[8–12]. Selectivity is an extremely important factor to decide applicability of a particular material for sensors[13–15]. Gas sensing materials can be made selectively for target gas by tuning its characteristics via doping or defect creation[16–20]. For development of a material as gas sensors, fundamental understanding of mechanisms responsible for gas adsorption is necessary. In this work, we have selected $NO_2$ as target gas to develop mechanism of gas adsorption on graphene surface. A comprehensive and deep understanding of $NO_2$ gas adsorption on defected and functionalized graphene can be useful in development of practical graphene-based gas sensors. Previous reports suggest that $NO_2$ gas adsorption can be enhanced through modifying graphene structure via creation of vacancies[21,22], doping impurities[22–26] and attaching chemical functional groups[27–30]. Using density functional theory (DFT), Lee *et al* elucidated that the monovacant graphene have strong $NO_2$ gas adsorption[21] and Zhou *et al* proposed that doping of transitional metal atoms (Cu, Ag and Au) can improve the sensing performance of graphene[23]. In an experimental study, Zhang *et al* demonstrated enhanced sensing performance after introducing single carbon vacancy defect into graphene[22]. Herein, we report an overall view on the effects of vacancies, adatom (Fe atom) and oxygenated groups on $NO_2$ gas adsorption behaviour of graphene surface using DFT.

The spin-polarized density functional theory calculations are carried out using SIESTA code based on numerical atomic orbitals basis set. The Perdew-Burke-Ernzerhof (PBE) generalized gradient approximation (GGA) is used as exchange-correlation energy. For all calculations, we have used 300 Ry as the plane wave cut-off energy and 7 x 7 x 1 as Monkhorst-Pack grid of **k** points. For Partial Density of State (PDOS) calculation, we applied a denser **k** point value that is 16 x 16 x 1. The structures can only be optimized when the energy difference between each element is less than $10^{-4}$ eV and the total force is less than 0.05eV/Å. A large 9 x 9 x 1 **k** points value was tested on pristine graphene structure to ensure that the 7 x 7 x 1 **k** points value is sufficient for the structure optimization process.

All basic graphene structures interacting with $NO_2$ are shown in **Figure 1**. In the present study, the modelled systems are based on a supercell of 6 x 5 graphene containing 60 carbon atoms. For a systematic study, we have used three type of structures- 1) Carbon vacancies varying from mono to tetra in graphene as shown in **Figure 1a**, 2) Single iron doped graphene structures as shown in **Figure 1b** and 3) Carboxyl and pyran group to understand the role of functional groups in graphene (**Figure 1c**). The $NO_2$ gas adsorption energy $E_a$ is calculated by following equation: $E_a = E_{substrate+NO_2} - E_{substrate} - E_{NO_2}$, where $E_{substrate+NO_2}$ and $E_{substrate}$ represent the energy of pristine or defective graphene after and before $NO_2$ gas adsorption. Here, $E_{NO_2}$ is the energy of single $NO_2$ gas molecule. The binding energy of Fe doping to graphene structure is calculated by following equation: $E_b = E_{Fe\text{-}doped\ graphene} - E_{vacant\ graphene} - E_{Fe}$. Where $E_{Fe\text{-}doped\ graphene}$ and $E_{vacant\ graphene}$ are the energy of vacant graphene after and before iron is introduced and $E_{Fe}$ is the energy of single iron atom.

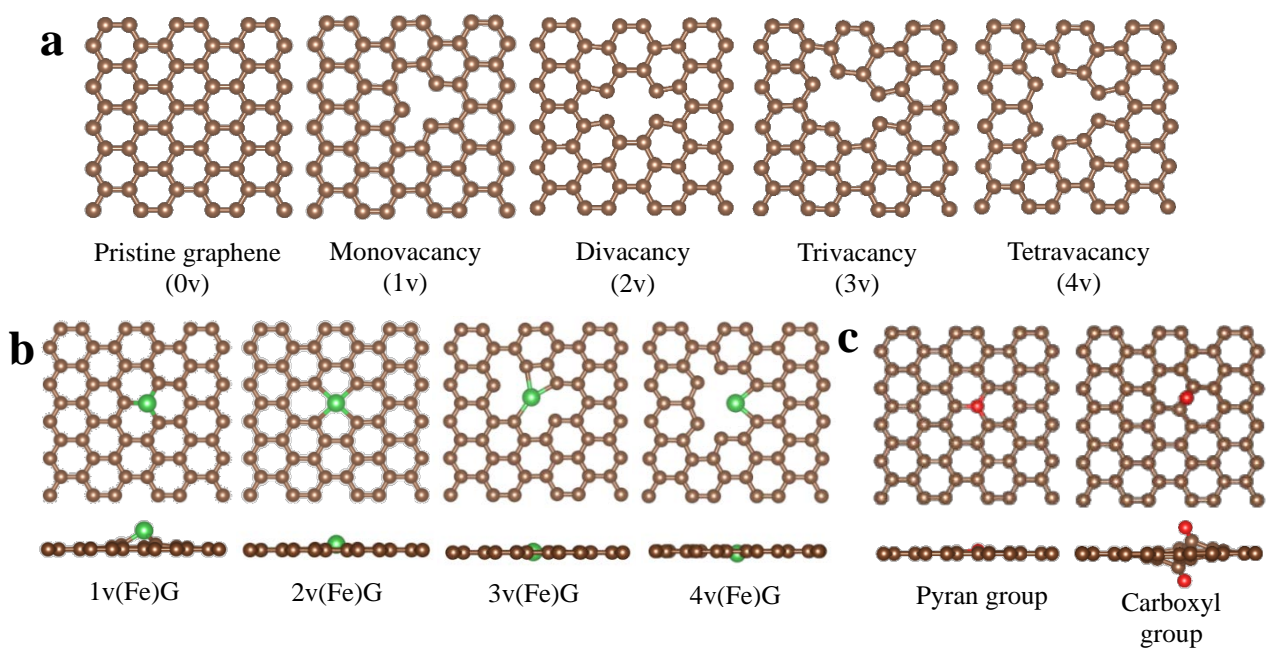

**Figure 1** Basic structure of graphenefor $NO_2$ gas adsorption. **a.** Carbon deficiency varying from monovacancy (1v) to tetravacancy (4v). **b**. Single Fe atom doping into the basic structure of **Figure 1a**. **c**. The main functional groups in reduced graphene oxide: pyran and carboxyl group. Brown colour is carbon; green colour is iron and red colour is oxygen.

## Results and Discussion
**Gas adsorption on pristine and carbon vacant graphene.** Based on the calculation of absorption energy for this system as depicted in **Figure 2a**, we can say that the carbon vacancies play an important role in enhancing $NO_2$ gas adsorption on graphene surface. For pristine graphene, the well-distributed charge on the iso-surface (**Figure 2b**) does not reveal any active sites for the $NO_2$ gas to interact, causing a physical adsorption type process with a small charge transfer of 0.003$e^-$ from $NO_2$ molecule to pristine graphene. Unsaturated carbon as exposed active sites can be found after introducing some carbon vacancies. These unsaturated carbon atoms in the vicinity of defect in the lattice apply a force to interact with $NO_2$ molecule causing an adsorption mechanism which is similar to a typical chemical adsorption process. Furthermore, the overlap of the charge density between $NO_2$ gas molecule and vacant graphene as shown in **Figure 2b** illustrates the formation of chemical bonding. This suggests that creation of vacancies causes a transition from physical adsorption to chemical adsorption on graphene. Moreover, we have observed that the structural configuration of $NO_2$ while interacting with graphene surface changes with creation of defects, which is clearly in **Figure 2b**. When $NO_2$ molecules approach the pristine graphene, it appears that the oxygen atoms prefer to interact with the graphene while in the case of graphene with defects, the excess electrons accumulated on the unsaturated carbon attract the unpaired electron in the nitrogen head of $NO_2$. Our study suggests that vacancies created by odd numbered carbon removal are more favourable for $NO_2$ - graphene surface interaction than the vacancies created by removing the even numbered carbon as shown in **Figure 2a**. This phenomenon can be attributed to presence of localized $sp^2$ dangling bond of carbon in odd numbered vacant type graphene.

We have studied the Partial Density of States (PDOS) to understand electronic density distribution for carbon atoms near and away from defect sites in carbon vacant graphene (**Figure 2c**). The monovacancy, bivacancy, trivacancy and tetravacancy are represented as 1vG, 2vG, 3vG and 4vG in this study. For monovacancy (**Figure 2c**), the carbon atom away from defects ( green circle) has steady electronic density suggesting no driving force induced for interacting with $NO_2$ gas molecules. The carbon atom adjacent to the single vacancy (blue circled) shows sharp PDOS peaks (**Figure 2c**) at far region above Fermi level and red circled carbon gives a peak at -0.74eV. The peak at -0.74eV in **Figure 2c** associated to red-circled carbon is derived from localized $sp^2$ dangling bonds[31] and thus suggesting a strong attraction to $NO_2$ gas molecules with a charge value of $0.029e^-$ that is transferred from 1vG to $NO_2$ molecule. With a observed magnetic moment of $0.88\mu_B$ which is close to the reported value of $1\mu_B$ [31], the red circled carbon is found to be the main contributor to the induced magnetism of monovacant graphene while interacting with $NO_2$ gas molecule [32,33].

In order to understand $NO_2$ adsorption for the trivacancy graphene (3vG) structure (odd numbered vacancy), we specifically focussed on the carbon atom near the defect sites (**Figure 2c**). Here, unlike the three carbon atoms shown in blue, green and pink circle, the red-circled carbon that shows sharp peaks near Fermi level can be the $NO_2$ adsorption site. Specifically, the peak above the Fermi level at 1.4eV for red-circled carbon reveals a strong force to capture the electrons from $NO_2$ molecules. Therefore, in 3vG the direction of charge transfer is from $NO_2$ gas molecule to the defective graphene surface unlike 1vG. Furthermore, it should bThe situations for 2vG and 4vG (even numbered) are quite different from the odd-numbered vacancies (1vG and 3vG) structures. For bivacancy structure (2vG in **Figure 2c**), the carbon away from the vacant site (pink colour) shows a flat PDOS curve suggesting a low likelihood for $NO_2$ to interact. Atoms near the vacant sites (red, blue and green colour), however provide a low peak at 2.5eV, still generate a weak force for binding $NO_2$ gas molecules. In the case of for 4vG, the PDOS peak at -0.8eV in both red and blue-circled atoms suggest tendency of these atoms to attract $NO_2$ molecules. On comparing the PDOS plots of $NO_2$ adsorption site in 2vG and 4vG structure, a relatively higher peak that is closer to the Fermi level indicates firm binding $NO_2$ molecules to 4vG graphene surface. Furthermore, the value of charge transfer for 2vG and 4vG are $0.017e^-$ and $0.012e^-$ given in **Figure 2b**e pointed out that in the $NO_2$ adsorption process, this red circled carbon atom in 3vG is the main contributor to the induced magnetic moment of $1.01\mu_B$, consistent with the reported value of $1.02\mu_B$[34].

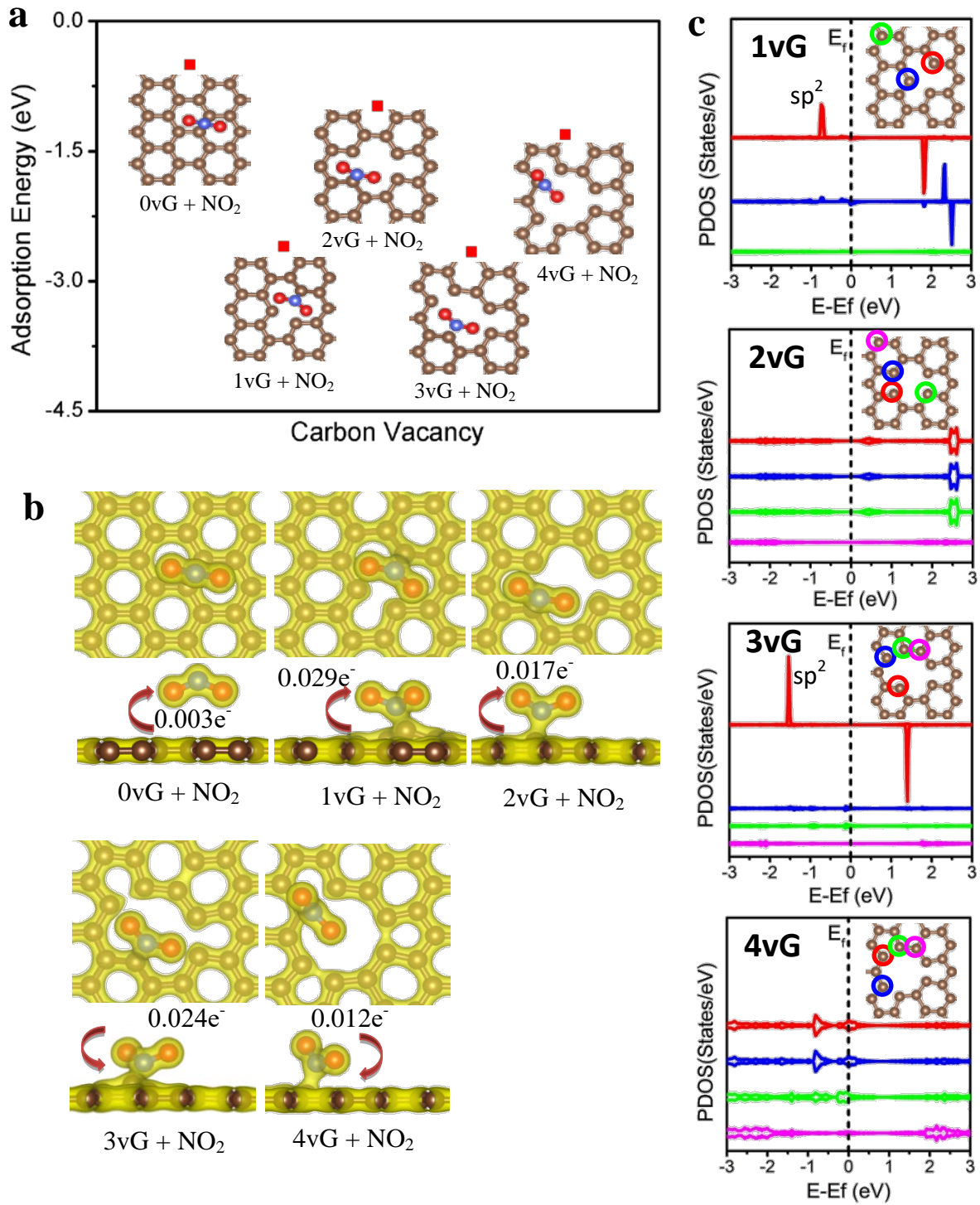

**Figure 2 a.** Adsorption energy variation with the number of carbon vacancies. **b.** The charge density isosurface (0, 1) and charge transfer of $NO_2$ adsorbed pristine and defective graphene (Brown is carbon, green is iron, purple is nitrogen and red is oxygen). **c.** Partial Density of States (PDOS) graph of carbon near vacancies. The colour for circled carbon atom is consistent with the colour in PDOS curve.

**Gas adsorption on Fe doped graphene.** Furthermore, the doping effect was studied by introducing single Fe atom into the mono-to- tetra vacant graphene. The Fe doped defected graphene structures are represented as 1v(Fe)G, 2v(Fe)G, 3v(Fe)G, and 4v(Fe)G. The binding energy for the Fe-doped defective graphene system is shown in **Figure 3a**. Our observations suggest that the Fe atom binds well in 1v, 2v and 3v graphene due to the availability of space provided for bond formation. As for tetravacancy, the excessive carbon removal makes a capacious room for Fe atom to be accommodated to form two bonds in 4v(Fe)G structure. The charge density graph shown in **Figure 3b** verifies the binding energy observation for Fe-graphene complex. The overlap of charge density between Fe and defective graphene reveals the bond formation therefore the number of Fe-C bond for 1v(Fe)G, 2v(Fe)G, 3v(Fe)G and 4v(Fe)G are 3, 4, 3 and 2 respectively. According to our study, the 2v(Fe)G shows the most stable structure due to the formation of four Fe-C bonds and with only two Fe-C bonds the 4v(Fe)G is the least stable structure. The direction of charge transfer from Fe (n-type dopant) to defected graphene is denoted as red and amounts of charge migration are in order of $0.075e^-$, $0.054e^-$, $0.053e^-$ and $0.333e^-$ for 1v(Fe)G-4(Fe)vG respectively in **Figure 3b**.

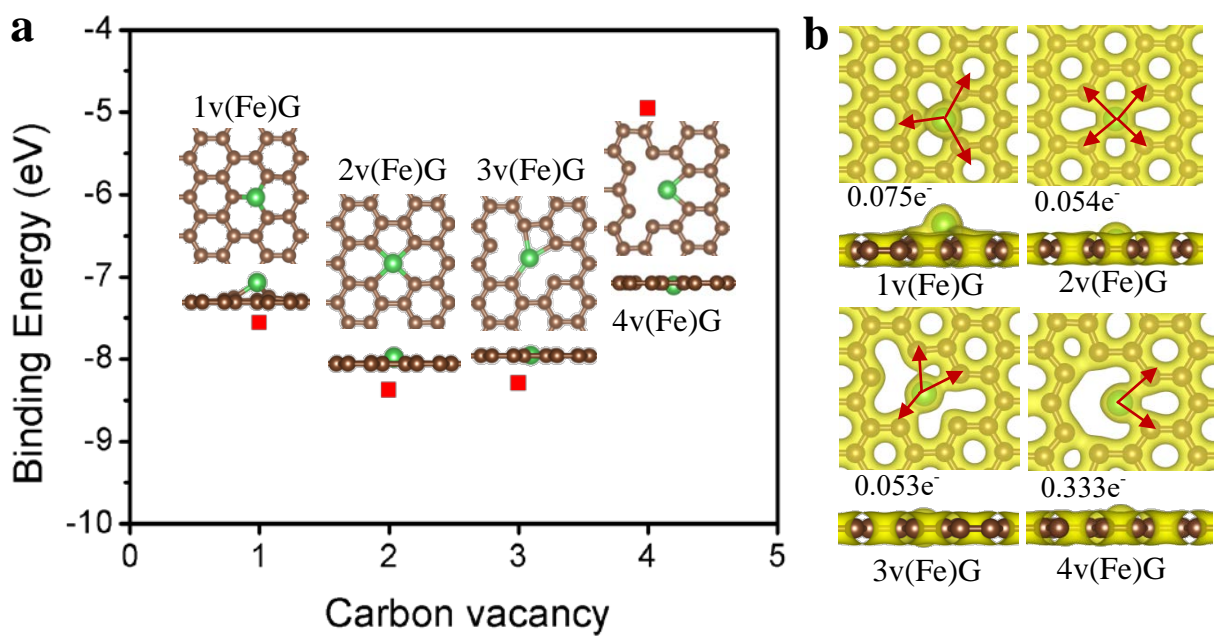

Figure 3 **a.** The binding energy of single Fe-doped graphene complex. **b.** The charge density isosurface (0, 1) and charge transfer (shown in the direction of red arrow) of single Fe-doped graphene complex.

We have found that the binding energy of 1v(Fe)G structure is lower than 2v(Fe)G structure which is not in agreement with the data reported Krasheninnikov *et al*[35]. However, our relaxed structures and associated characteristics are in well agreement. For example, the elevation height values ( distance of Fe atom outward defective graphene plane) of 1.28Å and 0.51Å for 1v(Fe)G and 2v(Fe)G respectively are in close match with corresponding value of 1.3Å and 0.55Å by Krasheninnikov *et al*[35]. Similarly the bond length (Fe-C) value of 1.8Å and 1.96Å for 1v(Fe )G and 2v(Fe)G respectively are close to the values 1.82Å and 1.98Å reported by Krasheninnikov *et al*[35]. $NO_2$ gas adsorption behavior on Fe-doped graphene sturcture is shown in **Figure 4**. The value of absorption energy shown in **Figure 4a** indicates 1v(Fe)G and 4v(Fe)G structure are bonded with $NO_2$ molecules more strongly than the 2v(Fe)G and 3v(Fe)G structures. Accoridng to the charge isosurface distribution shown in **Figure 4b**, the Fe-doped graphene are prone to interact with the oxygen in $NO_2$ molecule to form Fe-O bonds.

The extended charge distribution along the Fe-O bond for mono and tetra ( as shown in blue dash line in **Figure 4b)** suggest a greater bond strength that provides a higher $NO_2$ gas adsorption energy than the that of 2v(Fe)G and 3v(Fe)G.

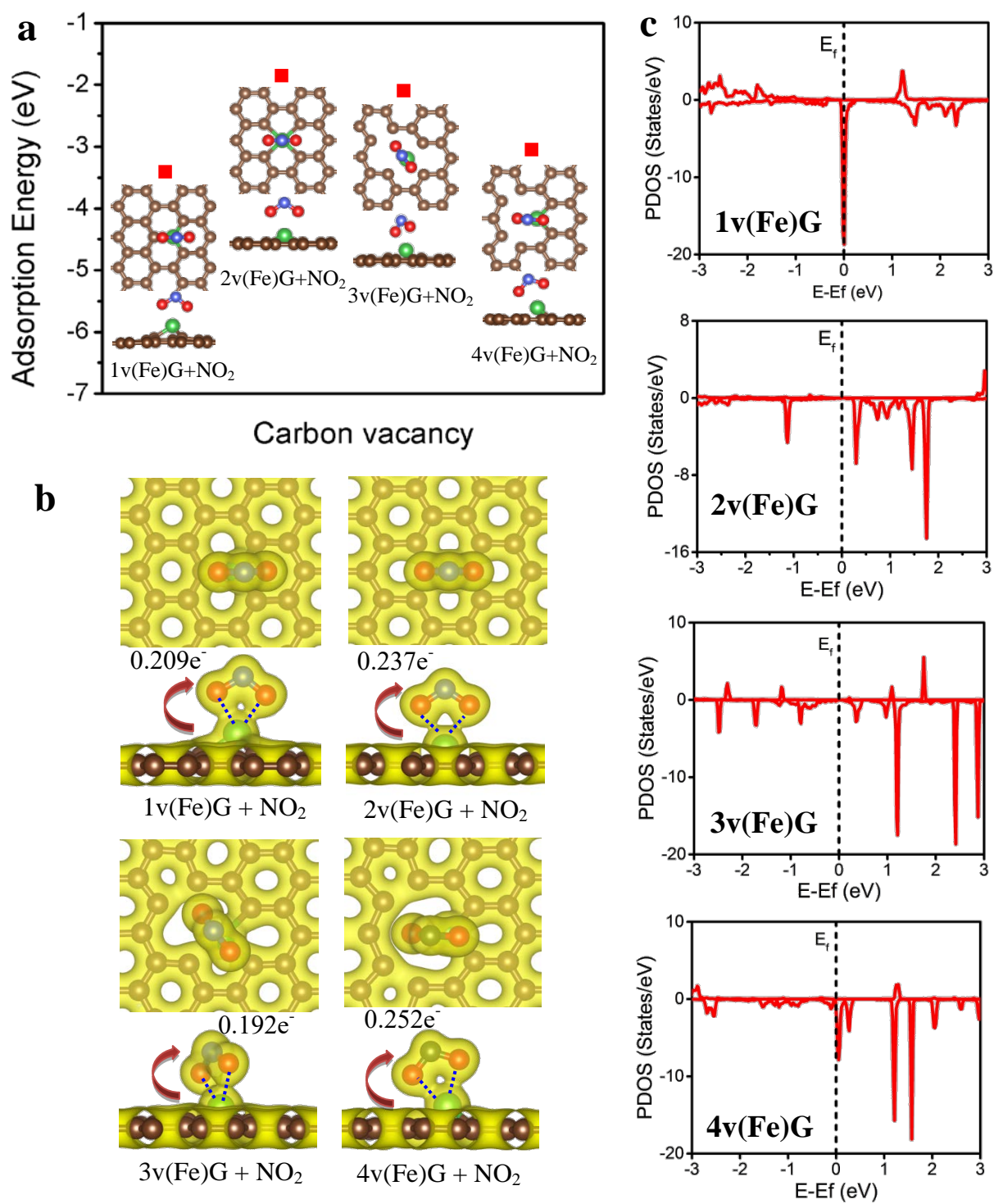

**Figure 4 a.** The $NO_2$ adsorption energy variation with the number of carbon vacancy dopped by Fe. **b.** The charge density isosurface (0, 1) and charge transfer of $NO_2$ adsorbed single Fe-doped graphene complex. **c.** Partial Density of States (PDOS) graph of 3d orbital of Fe atom before $NO_2$ gas adsorbed.

The Fe incorpotated defected graphene strucrure show better $NO_2$ gas adsorption behaviror than the only carbon deficiet defectced graphene sturcture this can be atributed to the excess free d electrons in iron. In order to understand this effect, we conducted PDOS mesurments of 3d orbital in Fe atom ( **Figure 4c).** In PDOS plots, a considerably higher peak closely located at the Fermi level in 1v(Fe)G and 4v(Fe)G indicates a significant attraction to capture $NO_2$. The PDOS plots for 2v(Fe)G and 3v(Fe)G structures also have sharp peaks but located aways from Fermi level suggesting a relatively low $NO_2$ adsorption energy. Moreover, **Figure 4b** presents the electrons are shifted from Fe-vacancy complex to $NO_2$ gas molecule with the value of $0.209e^-$, $0.237e^-$, $0.192e^-$ and $0.252e^-$ for 1-4 Fe-doped graphene structure.

**Gas adsorption on functionalised graphene.** We have also studied carboxyl and pyran group attached to graphene to understand the role of functional groups in gas adsorption process. Pyran and carbonyl group are two stable structural groups in the reduced graphene oxide[36] which is gas sensing material. According to our calculated values of adsorption energy (**Figure 5a)**, it appears that the pyran group facilitates the $NO_2$ adsorption more effectively than the carboxyl group. For pyran functionalized graphene, the unsaturated carbon atom near pyran group is tightly bonded to $NO_2$. The overlap of charge density between the $NO_2$ and graphene oxide **(Figure 5b)** as well as the large amount of charge ($0.084e^-$) reveal the formation of the C-N bonding. However, as no active sites can be found in carboxyl group, $NO_2$ gas can only be weakly adsorbed on the graphene oxide surface with a small value of charge ($0.01e^-$ ) transferred from $NO_2$ gas to graphene surface.

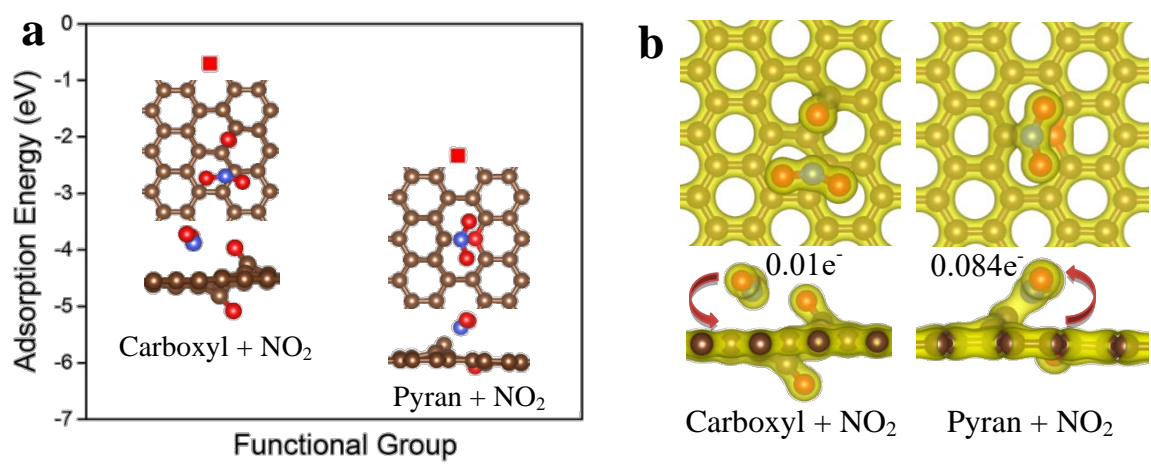

**Figure 5 a.** The $NO_2$ adsorption energy of pyran and carboxyl functional group in reduced graphene oxide. **b.** The charge density isosurface (0, 1) and charge transfer in $NO_2$ adsorbed in functional group.

Moreover, the ether group has been reported as an interesting ultrasensitive group for sensing $NO_2$ gas molecules with a reported binding energy value of $-0.212eV$[27]. This low $NO_2$ binding energy indicates a good reproducibility for developing the $NO_2$ gas sensor. In contrast, the strong binding of pyran and carbonyl group reported here is not fit for the $NO_2$ gas sensor but can provide the clue for the graphene-based catalyst or capture applications.

It can be concluded that the introduction of defects have an important role in $NO_2$ gas adsorption process. Single Fe dopant is most effective amongst other defected graphene studied for $NO_2$ gas enhancing $NO_2$ gas adsorption. In the case of oxygenated functional group modified graphene surface, the unsaturated carbon defect site near the pyran group is more attractive interaction

NO$_2$ than the carboxyl group and thus greatly enhances the gas adsorption characteristic. This study can be useful for future graphene-based adsorption applications as gas capture devices.